\documentclass[sigconf,screen]{acmart}

\AtBeginDocument{%
  }

\acmYear{2026}\copyrightyear{2026}
\setcopyright{cc}
\setcctype[4.0]{by}
\acmConference[FSE Companion '26]{34th ACM Joint European Software Engineering Conference and Symposium on the Foundations of Software Engineering}{July 5--9, 2026}{Montreal, QC, Canada}
\acmBooktitle{34th ACM Joint European Software Engineering Conference and Symposium on the Foundations of Software Engineering (FSE Companion '26), July 5--9, 2026, Montreal, QC, Canada}
\acmDOI{10.1145/3803437.3804875}
\acmISBN{979-8-4007-2636-1/26/07}

\usepackage{xspace}

\usepackage[utf8]{inputenc}
\usepackage[table]{xcolor}
\usepackage{xspace}
\usepackage{multirow}
\usepackage{subfig} 
\usepackage{balance} 
\usepackage{rotating}
\usepackage{enumitem}
\usepackage{listings}
\usepackage{lscape}
\usepackage[ruled,vlined]{algorithm2e}
\usepackage{tikz} 
\usepackage{enumitem}
\usepackage{listings}

\usetikzlibrary{positioning,shapes.geometric,shapes.misc}

\begin{document}

\title{From Helpful to Trustworthy: LLM Agents for Pair Programming }


\author{Ragib Shahariar Ayon}
\authornote{Advisor: Shibbir Ahmed, Assistant Professor, Texas State University, TX, USA.}
\affiliation{%
  \institution{Texas State University}
  \city{San Marcos}
  \state{TX}
  \country{USA}
}
\email{ipd21@txstate.edu}
\orcid{0009-0006-6372-5000}

\ccsdesc[500]{Software and its engineering~Formal software verification}

\keywords{Agentic AI, Large Language Model, Pair Programming}


\begin{abstract}
LLM-based coding agents are increasingly used to generate code, tests, and documentation. Still, their outputs can be plausible yet misaligned with developer intent and provide limited evidence for review in evolving projects. This limits our understanding of how to structure LLM pair-programming workflows so that artifacts remain reliable, auditable, and maintainable over time. To address this gap, this doctoral research proposes a systematic study of multi-agent LLM pair programming that externalizes intent and uses development tools for iterative validation. The plan includes three studies: translating informal problem statements into standards-aligned requirements and formal specifications; refining tests and implementations using automated feedback, such as solver-backed counterexamples; and supporting maintenance tasks, including refactoring, API migrations, and documentation updates, while preserving validated behavior. The expected outcome is a clearer understanding of when multi-agent workflows increase trust, along with practical guidance for building reliable programming assistants for real-world development.
\end{abstract}

\maketitle

\section{Introduction}
\label{sec:intro}
LLM-based coding agents are increasingly used for code generation, repair, and interactive assistance. Despite strong empirical performance, developers may still be cautious about relying on these tools as fully autonomous collaborators in realistic projects, since their outputs can appear plausible while remaining misaligned with developer intent and therefore still require human scrutiny~\cite{SWE-agent,survey,specRover,AutoCodeRover}. Improving trust in LLM-assisted development requires workflows that make intent explicit, enable systematic validation, and produce evolution-ready artifacts~\cite{roychoudhury2025agentic}.

We study a multi-agent system (MAS) in which a driver agent proposes artifacts and a navigator agent critiques them, each operating through role-specific prompts and separate interaction histories while sharing a persistent project context. Both roles are occupied by LLM agents, instantiated either as the same base model under different prompts or as models specialized for generation and review. We choose this driver-navigator design over a single-agent setup because it provides a minimal and interpretable separation between proposal and critique, and prior work suggests that such role specialization can improve code quality through structured, iterative feedback~\cite{zhang2024pair}. A natural concern with this setup is that, if the navigator is itself an LLM, the developer may simply be trading one auditability problem for another~\cite{SWE-agent,survey}. We address this by constraining the navigator to produce machine-checkable contracts and formal specifications rather than free-form judgments, which are then validated by deterministic verifiers that return proofs or counterexamples~\cite{endres2024can,specgen,ayon2026autore,ayon2026specpylot}. Trust thus shifts away from one model's assessment of another and toward externally verifiable evidence, leaving the developer with one concrete responsibility: confirming that the specification captures their intent.

Initial results show that verifier-guided LLM-based specification synthesis, through iterative feedback, improves correctness and completeness. Building on this, we plan to investigate pair-programming LLM agents as an end-to-end, evidence-driven workflow that translates intent into standards-aligned requirements and formal specifications, refines code and tests via automated feedback (e.g., SMT-based counterexamples), and sustains artifacts across software evolution.
\section{Background and Related Work}
\label{sec:background}
\textbf{Requirements and context generation:}
ArchCode extracts functional and non-functional requirements to guide code and test generation~\cite{han-etal-2024-archcode}, and SpecGen and AutoReSpec generate and refine verifiable specifications using verifier feedback~\cite{specgen,ayon2026autore}. These tools, however, target specification generation in isolation rather than embedding requirement elicitation within a pair-programming workflow. Our preliminary work begins to address this~\cite{ayon2026autore,ayon2026autojml}, but how a driver-navigator pair can co-produce standards-aligned requirements and formal specifications remains an open question.

\textbf{Code and test generation with automated feedback:}
AutoCodeRover and SpecRover couple iterative code search with patch generation and specification inference~\cite{AutoCodeRover,specRover}, Tian et al.\ improve equivalent mutant detection via code embeddings~\cite{SpecificationMisunderstandingTian}. These efforts operate in single-agent or single-stage settings and do not expose solver-backed counterexamples as auditable artifacts within a driver--navigator loop. Our proposed TDD study will examine whether such a setup improves consistency with requirements and produces more trustworthy, reproducible feedback.

\textbf{Maintenance and repository-scale evolution:}
Tufano et al.\ translate natural-language review comments into code transformations~\cite{codereviewtufano}, and Abreu et al.\ predict risky diffs to support stability-preserving gating at scale~\cite{releasedeploymentAbreu}. Neither uses pre-existing specifications and test suites as behavioral constraints to confirm that changes preserve validated behavior. Our proposed maintenance study will investigate whether anchoring refactoring, dependency upgrades, and documentation updates to such constraints reduces regressions and keeps generated artifacts trustworthy.
\section{Research progress and plan}
\paragraph{\textbf{Research Progress}}
In our initial study on formal specification generation, we introduced AutoReSpec, a verifier-guided collaborative LLM framework for JML synthesis that classifies programs by structural complexity, selects a pair of primary and fallback LLM pair, and iteratively refines specifications using verifier feedback; when the primary model fails, a collaborative fallback model receives the last failed candidate and verifier errors for focused recovery~\cite{ayon2026autore}. On a 72-program benchmark, AutoReSpec verified 67 programs, achieving a 58.2\% success probability and 69.2\% completeness, while reducing the average evaluation time by 26.89\% over prior methods.
Building on these results, we developed AutoJML to study the impact of ReAct-based LLM agents on specification generation. AutoJML automates JML specification synthesis via iterative verification and mutation-driven completeness feedback, and additionally leverages web-based context retrieval~\cite{ayon2026autojml}. On a 120-program benchmark, AutoJML verified 109 programs, achieving an average completeness of 79.3\%, with significant improvements over state-of-the-art baselines on challenging control-flow patterns, including multi-path loops (81.48\%) and nested loops (85.71\%).

\paragraph{\textbf{Future Plan}}
We want to investigate workflows in a pair programming, multi-agent setting, where driver and navigator LLMs use requirements and specifications to iteratively refine tests, implementations, and maintenance updates. Refinement is guided by automated feedback signals, including SMT-based counterexamples, while tests and specifications constrain changes to preserve validated behavior across refactoring, documentation, and upgrades. We study the following research question: How effective is the workflow at producing tests and implementations that are consistent with requirements and specifications? How do solver-backed counterexamples and automated feedback influence refinement behavior, and which types of programs benefit most? How does feedback affect trustworthiness, measured by pass rates, inconclusive outcomes, and reproducibility of failures? How well can agents perform maintenance tasks while preserving existing behavior? How effective are test and specification constraints at preventing regressions during iterative maintenance and migration? How accurate and useful are the generated documentation artifacts, and do they introduce misleading or biased statements?

\section{Conclusion}
\label{conclusion}
This doctoral research aims to shift multi-agent LLM pair programming from helpful assistance to trustworthy support in real-world software development. We will study a driver-and-navigator setup to quantify when multi-agent collaboration improves reliability and trust signals relative to single-agent baselines. It will investigate how agents can externalize intent as requirements, use tests and solver-backed feedback as auditable evidence during iteration, and maintain artifacts through refactoring, documentation, and evolving features at scale. The goal is to develop a practical workflow that produces reliable artifacts and trust signals to support safer adoption. in software engineering research and practice.

\balance
\bibliographystyle{ACM-Reference-Format}
\bibliography{biblography}

\end{document}